\documentclass[final,3p,times]{elsarticle}

\usepackage{geometry}
\usepackage{hyperref}
\usepackage{url}
\hypersetup{hypertex=true,colorlinks=true,linkcolor=blue,anchorcolor=blue,citecolor=blue}

\usepackage{amsmath}

\usepackage{listings}
\usepackage{xcolor}
\usepackage{ulem}

\definecolor{darkgreen}{rgb}{0,0.6,0}

\usepackage{amssymb}

\newcounter{bla}

\journal{Computer Physics Communications}

\begin{document}
	
	\begin{frontmatter}
		
		
		
		\title{TopoTB: A software package for calculating the electronic structure and topological properties of the tight-binding model}
		
		
		\author[a,b]{Xinliang Huang}
		\author[c,d]{Fawei Zheng \corref{author}}
		\author[a]{Ning Hao \corref{author}}
		
		\cortext[author] {Corresponding author.\\\textit{E-mail address:} fwzheng@bit.edu.cn (F. Zheng),  haon@hmfl.ac.cn (N. Hao)}
		\address[a]{Anhui Province Key Laboratory of Low-Energy Quantum Materials and Devices, High Magnetic Field Laboratory, HFIPS, Chinese Academy of Sciences, Hefei, Anhui 230031, China}
		\address[b]{Science Island Branch of Graduate School, University of Science and Technology of China, Hefei, Anhui 230026, China}
		\address[c]{Centre for Quantum Physics, Key Laboratory of Advanced Optoelectronic Quantum Architecture and Measurement (MOE), School of Physics, Beijing Institute of Technology, Beijing 100081, China}
		\address[d]{Beijing Key Lab of Nanophotonics $\&$ Ultrafine Optoelectronic Systems, School of Physics, Beijing Institute of Technology, Beijing 100081, China}
		
		\begin{abstract}
			
			We present TopoTB, a software package written in the Mathematica language, designed to compute electronic structures, topological properties, and phase diagrams based on tight-binding models. TopoTB is user-friendly, with an interactive user interface that enables the tuning of model parameters for fitting the target energy bands in a WYSIWYG way. In addition, TopoTB also includes functionalities for processing results from Density Functional Theory calculations. The outputs of TopoTB are rich and readable, and they can be displayed in various styles. These features make TopoTB a useful tool for the theoretical study of materials.
			
		\end{abstract}
		
		
	\end{frontmatter}
	
	
	
	{\bf PROGRAM SUMMARY}
	
	\begin{small}
		\noindent
		{\em Program Title:} TopoTB                                       \\
		{\em CPC Library link to program files:} (to be added by Technical Editor) \\
		{\em Developer's repository link:} https://github.com/xlhuang-phy/TopoTB \\
		{\em Code Ocean capsule:} (to be added by Technical Editor)\\
		{\em Licensing provisions:} GPLv3  \\
		{\em Programming language:} Mathematica                                  \\
		{\em Nature of problem:}
        Analyze the electronic energy bands of DFT calculations, calculate the electronic structure and topological properties of the tight-binding model (including the manually set models and the one obtained from other software). \\
		{\em Solution method:} 
		The Wilson loop method was used to calculate the Berry phase, Berry curvature, Chern number, and $\mathbb{Z}_2$ number. In addition, the Shiozaki method was also implemented to calculate the $\mathbb{Z}_2$ number. \\
		\\
		
	\end{small}

	\section{Introduction}
	\label{Introduction}
	
	Nontrivial topological states have always been a hot research topic in condensed matter physics, and related materials, their theoretical models have also been attracting much attention \cite{moore2010birth, hasan2010colloquium, qi2011topological, bansil2016colloquium, lv2021experimental, xiao2021first}. For instance, the quantum spin Hall effect in two-dimensional HgTe/CdTe quantum wells \cite{bernevig2006quantum}, the Dirac-type dispersion of surface states in three-dimensional strong topological insulator Bi$_2$Se$_3$ \cite{zhang2009topological, xia2009observation}, and so on. These developments often progress in tandem with experiments, first-principles calculations, and theoretical studies. In experiments, the topological properties are determined by measuring edge states or surface states using angle-resolved photoemission spectroscopy (ARPES). In first principles calculations, electronic structure calculations can be performed using DFT code software such as Vienna ab initio simulations package (VASP) \cite{PhysRevB.54.11169, KRESSE199615}, Quantum-Espresso \cite{giannozzi2009quantum}, OpenMX \cite{openmx}, ABACUS \cite{li2016large}, etc. Then, topological properties can be calculated through the software's built-in scripts or interfaces with other software. In theoretical studies, for example, the tight-binding (TB) model, the existence of topology is determined by calculating the Berry phase \cite{PhysRevLett.62.2747, RevModPhys.82.1959}, Berry curvature \cite{PhysRevLett.49.405, Chang_2008}, Chern number \cite{PhysRevLett.49.405}, $\mathbb{Z}_2$ number \cite{PhysRevLett.95.146802}, etc. These topological related physical quantities require eigenstates, i.e., wave functions, which are not unique in quantum mechanics, posing great difficulties for numerical calculations.
	
	There are many software packages available for calculating topological properties. For example, WannierTools \cite{WU2018405} is a open-source software package that requires \lstinline{wannier90_hr.dat} files generated by Wannier90 \cite{MOSTOFI2008685}, and provides powerful tools and support for studying the topological properties and phase transitions of materials. Z2pack \cite{PhysRevB.95.075146} is a Python software package that generates the TB models and calculates related topological properties through the Wannier charge centers (WCCs). PYATB \cite{JIN2023108844} is an efficient Python package for computing electronic structures, topological properties, and optical properties using ab initio TB model, serving as a post-processing tool for ABACUS. These software packages, as well as other related software packages \cite{pythtb, NAKHAEE2020107379, LI2023108632, SAINI2022108147, tyner2023berryeasy, FENG20121849}, are excellent tools in specific cases.
	
	Here, we introduce the TopoTB software package, which employs Wilson loop methods and Shiozaki methods without the need for any gauge fixing to compute the topological numbers of the TB model, effectively circumventing the ambiguity of wave functions. Additionally, our software package supports the computation of electronic structures for the $\mathbf{k} \cdot \mathbf{p}$ model. Compared to other software packages such as WannierTools, Z2pack, and PYATB, our software package can directly input the model Hamiltonian matrix and corresponding lattice vectors to compute electronic structures, such as band structures, density of states, spin textures, and more. For computations involving topological properties like Berry phase, Berry curvature, Chern number, $\mathbb{Z}_2$ number, etc., our software package requires supplementary input, such as the number of occupied states. These are the basic capabilities of the software package. In addition, our software package allows for interactive operation of the band structure, facilitating convenient observation of the changes in the bands. Particularly, by utilizing the band structure from DFT codes calculations as a reference, our software can be used to tune the parameters in the model. We also plan to develop the interfaces with other softwares, such as Wannier90, WannierTools, etc. For the construction of the model, the MagneticTB \cite{ZHANG2022108153} and MagneticKP \cite{zhang2023magnetickp} software packages may be helpful. So, the TopoTB software package focuses on handling the electronic structure and topological properties of the TB model, and the interface is quite user-friendly.
	
	The organizational structure of this paper is as follows. In Section \ref{Methods}, we introduce the basic theories related to software packages. In Section \ref{Capabilities}, we introduce the capabilities of software packages. In Section \ref{Installation}, we introduce the installation and usage of software packages. In Section \ref{Examples}, we provide relevant examples of the main capabilities of the software package. In Section \ref{Conclusion}, we present the conclusion.
	
	\section{Methods}
	\label{Methods}
	\subsection{TB model}
	
	The TB model, constructed upon the tight-binding approximation, employs linear combination of atom orbitals (LCAO) as its basis functions. It establishes a direct relationship between the electronic band structure of a crystal and the electronic energy levels of the constituent atoms in their isolated states. For the TB model, there are two gauges called ``lattice gauge" and ``atomic gauge". To describe these gauges, we consider $N$ lattices, each with $M$ atomic orbitals. If we denote the creation and annihilation operators on the $\alpha$-th atomic orbital in the $i$-th lattice as $c_{i\alpha}^{\dagger}$ and $c_{i\alpha}$ ($i=1,2,...,N; \alpha=1,2,...,M$), respectively, then the Hamiltonian can be written as
	\begin{align}
		H=\sum_{i, j}\sum_{\alpha, \beta} t_{\alpha \beta}^{ij} c_{i\alpha}^{\dagger} c_{j\beta}
	\end{align}
	where $t_{\alpha \beta}^{ij}=\langle i\alpha |H| j\beta \rangle $. It represents the hopping matrix element from the $\alpha$-th atomic orbital in the $i$-th lattice to the $\beta$-th atomic orbital in the $j$-th lattice. 
	This Hamiltonian is in real space, but our focus is in momentum space. Therefore, we need to perform a Fourier transform. It's important to note that different Fourier transforms will lead to different gauges for the TB models. 
	
\begin{figure}
		\centering
		\includegraphics[scale=0.85]{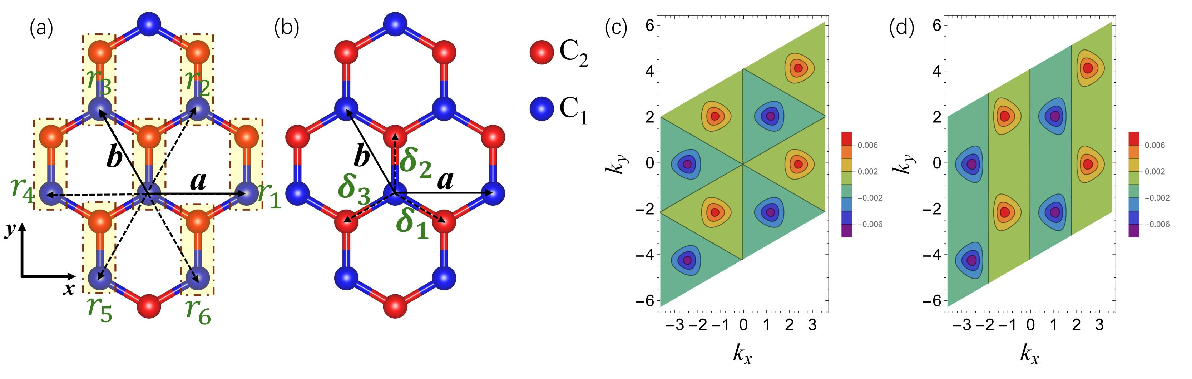}
		\caption{
			(a) A schematic diagram of the hopping of the ``lattice gauge". In the figure, $\mathbf{a}$ and $\mathbf{b}$ represent lattice vectors, dashed boxes represent a lattice (or primitive cell), and $\mathbf{r}_1$ to $\mathbf{r}_6$ represent hopping vectors from the central lattice to the six nearest neighboring lattices. (b) A schematic diagram of the hopping of the ``atomic gauge". In the figure, $\mathbf{a}$ and $\mathbf{b}$ represent lattice vectors, $\boldsymbol{\delta}_1$, $\boldsymbol{\delta}_2$, and $\boldsymbol{\delta}_3$ represent the three nearest neighbor hoppings from $C_1$ to $C_2$, and the three nearest neighbor hoppings from $C_2$ to $C_1$ are represented by their opposite vectors. 
			(c) and (d) are the Berry curvatures of the TB model of graphene constructed in (a) and (b), respectively.
		}
		\label{fig1}
	\end{figure}

	Under the ``lattice gauge", the atomic orbital information inside the lattice is ignored implicitly, so the Hamiltonian of the momentum space is written as
	\begin{align}\label{eq2}
		H(\mathbf{k})=\sum_{i, j} \left[ \sum_{\alpha, \beta} t_{\alpha \beta}^{ij} c_{i\alpha}^{\dagger} c_{j\beta} \right] e^{i \mathbf{k}\cdot \mathbf{R}_{ij}}
	\end{align}
	where $\mathbf{R}_{ij}=\mathbf{R}_{j}-\mathbf{R}_{i}$, and $\mathbf{R}_i$ is the lattice vector. That is to say, when constructing the TB model, we need to consider intracell and intercell hoppings separately, and the Fourier transform is performed on these two types of hoppings. Therefore, the Hamiltonian satisfies $H(\mathbf{k}+\mathbf{G})=H(\mathbf{k})$, where $\mathbf{G}$ is the reciprocal lattice vector, which is in Bloch form. A schematic diagram of this specification is shown in Fig.\ref{fig1}(a). For this  hexagonal honeycomb lattice, we consider nearest neighbor hoppings and obtain
	\begin{equation}\label{eq3}
		\begin{split}
			H(\mathbf{k}) &= \begin{pmatrix}
				M & 0 \\
				0 & -M
			\end{pmatrix} + 
			\begin{pmatrix}
				0 & t \\
				t & 0
			\end{pmatrix} \\
			&+ \begin{pmatrix}
				0 & 0 \\
				0 & 0  
			\end{pmatrix} e^{i \mathbf{k}\cdot \mathbf{r}_1} +
			\begin{pmatrix}
				0 & t \\
				0 & 0  
			\end{pmatrix} e^{i \mathbf{k}\cdot \mathbf{r}_2}+
			\begin{pmatrix}
				0 & t \\
				0 & 0  
			\end{pmatrix} e^{i \mathbf{k}\cdot \mathbf{r}_3} \\
			&+ \begin{pmatrix}
				0 & 0 \\
				0 & 0  
			\end{pmatrix} e^{i \mathbf{k}\cdot \mathbf{r}_4}+
			\begin{pmatrix}
				0 & 0 \\
				t & 0  
			\end{pmatrix} e^{i \mathbf{k}\cdot \mathbf{r}_5}+
			\begin{pmatrix}
				0 & 0 \\
				t & 0  
			\end{pmatrix} e^{i \mathbf{k}\cdot \mathbf{r}_6} \\
			&=\begin{pmatrix}
				M & t+t e^{i \mathbf{k}\cdot (\mathbf{a}+\mathbf{b})}+t e^{i \mathbf{k}\cdot \mathbf{b}} \\
				t+t e^{-i \mathbf{k}\cdot (\mathbf{a}+\mathbf{b})}+t e^{-i \mathbf{k}\cdot \mathbf{b}} & -M
			\end{pmatrix}
		\end{split}
	\end{equation}
	where $M$ is the potential energy at $C_1$ and $C_2$ sites, $t$ is the hopping parameter, $\mathbf{r}_1$ to $\mathbf{r}_6$ represent hopping vectors from the central lattice to the six nearest neighbor lattices. It can be seen from this that if a hopping parameter in the TB model is independent of $\mathbf{k}$, then the TB model is ``lattice gauge". The symmetry of the Berry curvature obtained in this gauge is broken, as shown in Fig.\ref{fig1}(c). The calculation of Berry curvature can be found in the following text.
		
	In the ``atomic gauge", we consider the hoppings of each atomic orbital in the lattice, and the Hamiltonian can be written as
	\begin{align}\label{eq4}
		H(\mathbf{k})=\sum_{i, j} \sum_{\alpha, \beta} \left[ t_{\alpha \beta}^{ij} c_{i\alpha}^{\dagger} c_{j\beta} e^{i \mathbf{k}\cdot (\mathbf{R}_{ij}+\mathbf{d}_{\alpha \beta})} \right]
	\end{align}
	where $\mathbf{R}_{ij}=\mathbf{R}_{j}-\mathbf{R}_{i}$, and $\mathbf{R}_i$ is the lattice vector; $\mathbf{d}_{\alpha \beta}=\mathbf{d}_{\beta}-\mathbf{d}_{\alpha}$, and $\mathbf{d}_\alpha$ is the position of atoms in a lattice. Applying this, Fig.\ref{fig1}(b) shows the ``atomic gauge", and we can easily obtain the Hamiltonian that takes into account the three nearest neighbor hopping, written as
	\begin{equation}\label{eq5}
		\begin{split}
			H(\mathbf{k}) &= \begin{pmatrix}
				M & 0 \\
				0 & -M
			\end{pmatrix} +
			\begin{pmatrix}
				0 & t e^{i \mathbf{k}\cdot \boldsymbol{\delta}_1} \\
				t e^{-i \mathbf{k}\cdot \boldsymbol{\delta}_1} & 0
			\end{pmatrix} + 
			\begin{pmatrix}
				0 & t e^{i \mathbf{k}\cdot \boldsymbol{\delta}_2} \\
				t e^{-i \mathbf{k}\cdot \boldsymbol{\delta}_2} & 0
			\end{pmatrix} +
			\begin{pmatrix}
				0 & t e^{i \mathbf{k}\cdot \boldsymbol{\delta}_3} \\
				t e^{-i \mathbf{k}\cdot \boldsymbol{\delta}_3} & 0
			\end{pmatrix} \\
			&= \begin{pmatrix}
				M & t e^{i \mathbf{k}\cdot \boldsymbol{\delta}_1}+t e^{i \mathbf{k}\cdot \boldsymbol{\delta}_2}+t e^{i \mathbf{k}\cdot \boldsymbol{\delta}_3} \\
				t e^{-i \mathbf{k}\cdot \boldsymbol{\delta}_1}+t e^{-i \mathbf{k}\cdot \boldsymbol{\delta}_2}+t e^{-i \mathbf{k}\cdot \boldsymbol{\delta}_3} & -M
			\end{pmatrix}
		\end{split}
	\end{equation}
	where $M$ is the potential energy at $C_1$ and $C_2$ sites, $t$ is the hopping parameter, $\boldsymbol{\delta}_1$, $\boldsymbol{\delta}_2$, and $\boldsymbol{\delta}_3$ represent the three nearest neighbor hoppings from $C_1$ to $C_2$, and the three nearest neighbor hoppings from $C_2$ to $C_1$ are represented by their opposite vectors. Because of $\mathbf{G}\cdot\boldsymbol{\delta}_i \neq 2\pi n$, where $\mathbf{G}$ is the reciprocal lattice vector and $n$ is an integer, this gauge is not in Bloch form. Under this gauge, the ``environment" around atomic orbitals is well defined, so the symmetry of Berry curvature is protected, as shown in Fig.\ref{fig1}(d). The calculation of Berry curvature can be found in the following text.
	
	Regarding these two gauges, they correspond to Equation(\ref{eq2}) and Equation(\ref{eq4}), respectively. For Equation(\ref{eq2}), due to $H(\mathbf{k}+\mathbf{G})=H(\mathbf{k})$, its eigenstate satisfies the Bloch form, and the Bloch wave function will be obtained, which is
	\begin{align}\label{eq6}
		| \psi_{\mathbf{k}}(\mathbf{r}) \rangle=e^{i \mathbf{k}\cdot\mathbf{r}} | u_{\mathbf{k}}(\mathbf{r}) \rangle
	\end{align}
	where $u_{\mathbf{k}}(\mathbf{r}+\mathbf{R}_n)= u_{\mathbf{k}}(\mathbf{r})$. On the contrary, the eigenstate of Equation(\ref{eq4}) is
	\begin{align}\label{eq7}
		| u_{\mathbf{k}}(\mathbf{r}) \rangle = | \psi_{\mathbf{k}}(\mathbf{r}) \rangle e^{-i \mathbf{k}\cdot\mathbf{r}}
	\end{align}
	The result is the periodic part of the Bloch wave function. Equation(\ref{eq6}) and Equation(\ref{eq7}) are also mentioned in Wanniertools \cite{WU2018405}. We have provided two different derivation methods above, corresponding to ``lattice gauge" and ``atomic gauge", respectively. For these two gauges, we can also convert them through a gauge transformation \cite{bernevig2013topological}. The formula for transformation \cite{yusufaly2013tight, pythtb} is
	\begin{align}
		H_{lattice}(\mathbf{k})= e^{i \mathbf{k} \cdot \mathbf{d}_{\beta \alpha}} H_{atomic}(\mathbf{k})
	\end{align}
	where $\mathbf{d}_{\beta \alpha}=\mathbf{d}_\alpha-\mathbf{d}_\beta$. We need to pay attention to the gauges used for each TB model. The two gauges will not affect the properties obtained by considering the entire Brillouin zone. But for the local properties of the Brillouin zone, such as Berry phase, Berry curvature, etc., we should choose ``atomic gauge".
	
	From the above content, we can see that these two gauges have an impact on the eigenstates (i.e. wave functions). The calculation of topological numbers requires wave function information, so we should pay attention to these two gauges. However, for electronic structure calculations, such as energy bands, density of states (DOS), Fermi surfaces, etc., only eigenvalue information is often required. The eigenvalues of these two gauges are exactly the same, so we don't need to distinguish them.
	
	In the TB model, in addition to paying attention to gauges, the basis of the Hamiltonian also needs special attention. For models considering spin-orbit coupling (SOC), common bases include ``orbital grouping" and ``spin grouping". For ``orbital grouping", the orbital space is placed at the front and the spin space is placed at the back, with the basis represented as $(\uparrow, \downarrow, \uparrow, \downarrow, ...)$, where $\uparrow, \downarrow$ is the spin index, the orbital index is ignored. For ``spin grouping", the spin space is placed at the front and the orbital space is placed at the back, with the basis represented as $(\uparrow, \uparrow, ..., \downarrow, \downarrow, ...)$, where $\uparrow, \downarrow$ is the spin index, the orbital index is ignored.
	
	\subsection{Spin texture}
	
	In two-dimensional systems with inversion symmetry breaking, there exists a potential gradient at the surface or interface, and the SOC of electrons leads to band splitting, resulting in a spin-momentum locked spin texture, which is called the Rashba effect \cite{YuABychkov1984, WOS:000360192000018}. To generate this effect, Rashba SOC needs to be considered in the TB model, written as
	\begin{align}
		H_{Rashba}=i \lambda_R \sum_{i,j} c_i^{\dagger} (\mathbf{s} \times \hat{\mathbf{d}}_{ij})_z c_j
	\end{align}
	where $c_i(c_i^{\dagger})$ is the annihilation (creation) operator at site $i$, $\mathbf{s}=(s_x, s_y, s_z)$ is the Pauli matrix, and $\hat{\mathbf{d}}_{ij}$ is the unit vector between lattice points, so there is
	\begin{align}
		\mathbf{s} \times \hat{\mathbf{d}}_{ij} = s_x \hat{d}_{y} - s_y \hat{d}_{x}
	\end{align}
	After adding Rashba SOC to the TB model, the spin texture can be calculated as follows
	\begin{align}
		S_i(\mathbf{k}) = \langle \psi(\mathbf{k}) | \Omega_i | \psi(\mathbf{k}) \rangle
	\end{align}
	where $\Omega_i=\tau_0 \otimes s_i$ (``orbital grouping") or $\Omega_i=s_i \otimes \tau_0$ (``spin grouping"), and $i=x, y, z$ is the $x$, $y$, and $z$ components of the Pauli matrix. The $\tau_0$ and $s_i$ represent orbital space and spin space, respectively.
	
	\subsection{Chern number calculation}
	\subsubsection{Berry phase and Berry curvature}
	
	The Berry phase \cite{PhysRevLett.62.2747, RevModPhys.82.1959}, also known as geometric phase, is measurable and depends on the chosen path. The path integral of a closed path in a parameter space is the Berry phase, where the parameter space is the reciprocal space and the Berry phase is written as
	\begin{align}
		\phi_n = \oint_{\mathcal{C}} d\mathbf{k} \cdot \mathbf{A}_n(\mathbf{k})
	\end{align}
	where $n$ is the band index, $\mathbf{A}_n(\mathbf{k})$ is the Berry connection, defined as,
	\begin{equation}
		\begin{split}
			\mathbf{A}_n(\mathbf{k}) &= i \langle u_n(\mathbf{k}) | \nabla_{\mathbf{k}} | u_n(\mathbf{k}) \rangle \\
			&=- \operatorname{Im} \left[ \langle u_n(\mathbf{k}) | \nabla_{\mathbf{k}} | u_n(\mathbf{k}) \rangle \right]
		\end{split}
	\end{equation}
	Analogous to the formula $\mathbf{B}=\nabla \times \mathbf{A}$ in electromagnetics, we can define the Berry curvature \cite{PhysRevLett.49.405, Chang_2008}, written as
	\begin{align}
		\mathbf{\Omega}^n(\mathbf{k})=\nabla_{\mathbf{k}} \times \mathbf{A}_n(\mathbf{k})
	\end{align}
	The Berry curvature has different forms and can be written as
	\begin{equation}
		\begin{split}
			\mathbf{\Omega}_{\mu\nu}^n(\mathbf{k}) &= \partial_\mu (\mathbf{A}_n(\mathbf{k}))_\nu -\partial_\nu (\mathbf{A}_n(\mathbf{k}))_\mu \\
			&= i \left[ \langle \partial_\mu u_n(\mathbf{k}) | \partial_\nu u_n(\mathbf{k}) \rangle  -  \langle \partial_\nu u_n(\mathbf{k}) | \partial_\mu u_n(\mathbf{k}) \rangle  \right] \\
			&= -2 \operatorname{Im} \left[ \langle \partial_\mu u_n(\mathbf{k}) | \partial_\nu u_n(\mathbf{k}) \rangle \right]
		\end{split}
	\end{equation}
	It can also be written using the Kubo formula as
	\begin{equation}
		\begin{split}
			\mathbf{\Omega}_{\mu\nu}^n(\mathbf{k}) &= - \operatorname{Im} \sum_{m \neq n} \frac{\langle u_n(\mathbf{k}) | \nabla_{\mathbf{k}}H(\mathbf{k}) | u_m(\mathbf{k}) \rangle \times \langle u_m(\mathbf{k}) | \nabla_{\mathbf{k}}H(\mathbf{k}) | u_n(\mathbf{k}) \rangle}{(E_n-E_m)^2} \\
			&= i \sum_{m \neq n} \frac{\langle u_n(\mathbf{k}) | \partial_\mu H(\mathbf{k}) | u_m(\mathbf{k}) \rangle \langle u_m(\mathbf{k}) | \partial_\nu H(\mathbf{k}) | u_n(\mathbf{k}) \rangle  -  (\mu \leftrightarrow \nu)}{(E_n-E_m)^2}
		\end{split}
	\end{equation}
	where $\partial_\mu=\partial_{{k}_\mu}$ and $\partial_\nu=\partial_{{k}_\nu}$. These formulas are for single-band cases. For multi-band cases, there may be crossing or degeneracy of bands, which is non Abelian. At this point, the formulas for calculating Berry phase and Berry curvature are written as follows
	\begin{gather}
		\phi = \oint_{\mathcal{C}} d\mathbf{k} \cdot \operatorname{Tr}[\mathbf{A}_{nm}(\mathbf{k})] \\
		\mathbf{\Omega}_{\mu\nu}(\mathbf{k}) = \operatorname{Tr}[-2 \operatorname{Im} \left[ \langle \partial_\mu u_n(\mathbf{k}) | \partial_\nu u_m(\mathbf{k}) \rangle \right] ]
	\end{gather}
	where $\mathbf{A}_{nm}(\mathbf{k})=i \langle u_n(\mathbf{k}) | \nabla_{\mathbf{k}} | u_m(\mathbf{k}) \rangle$, $n(m)$ is the band index. The calculation formulas for Berry phase and Berry curvature given above involve taking the derivative of the wave function. Due to the non uniqueness of the wave function, which can differ by one phase factor, it poses difficulties in numerical calculations. In summary, if the TB model is simple, you can apply formulas for analytical calculations, but numerical calculations need to be very careful.
	
	\subsubsection{Chern number}
	
	Integrating the Berry curvature in the first Brillouin zone (BZ) yields the Chern number, also known as the Thouless-Kohmoto-Nightingale-Nijs (TKNN) number \cite{PhysRevLett.49.405}, which is defined as
	\begin{align}
		C_n=\frac{1}{2\pi} \int_{BZ} d\mathbf{k} \cdot \mathbf{\Omega}_{k_x,k_y}^n(\mathbf{k})
	\end{align}
	If we directly use the definition of Berry curvature to calculate the Chern number, it will also bring great difficulties to numerical calculations. Therefore, we choose the Fukui-Hatsugai-Suzuki method \cite{doi:10.1143/JPSJ.74.1674} to calculate the Chern number. This method is also known as the Wilson loop method \cite{WOS:000809131400001, asboth2016short, Topology} for calculating Chern number. The formula for calculating lattice field strength on a discrete grid is
	\begin{equation}\label{eq20}
		\begin{split}
			\tilde{F}_l&= -\operatorname{Im} \left[ \operatorname{ln} \left[ \prod_p \operatorname{Det}\left[\langle u_n({\bf{k}}_{l,p})|u_{m}({\bf{k}}_{l,p+1})\rangle \right] \right] \right] \\
			&=-\operatorname{Arg} \left[\prod_p \operatorname{Det} \left[\langle u_n({\bf{k}}_{l,p})|u_{m}({\bf{k}}_{l,p+1})\rangle \right] \right]
		\end{split}
	\end{equation}
	where $l$ represents the BZ grid into l small plaquettes, $n(m)$ labels the band index, and ${\bf{k}}_{l,p}$ are momenta on the corners of the lattice. By summing the lattice field strength in momentum space, we can obtain the Chern number, as follows
	\begin{align} \label{eq19}
		C = \frac{1}{2 \pi} \sum_l \tilde{F}_l
	\end{align}
	where $l \in [0, 2 \pi]\times [0, 2\pi]$. Note that this formula can calculate the Chern number of multi-band case. The Hall conductance is the total number of occupied states expressed as $\sigma_{xy}=\frac{e^2}{h} \sum_i C_i$, where $i$ represents the band index of the occupied states.
	
	When l is large enough, Equation(\ref{eq20}) is the Berry curvature at the l-th small plaquettes, i.e., $\Omega_l=\tilde{F}_l$. For the Berry phase, we need to calculate along a loop, so the Berry phase is defined as
	\begin{equation}
		\begin{split}
			\phi = -\operatorname{Im} \left[ \operatorname{ln} \left[ \prod_p \operatorname{Det} \left[ M^p \right] \right] \right] =-\operatorname{Arg} \left[\prod_p \operatorname{Det}\left[ M^p \right] \right]
		\end{split}
	\end{equation}
	where $M_{nm}^p=\langle u_n({\bf{k}}_{p})|u_{m}({\bf{k}}_{p+1})\rangle$, $p$ is on the selected loop.
	
	\subsection{\texorpdfstring{$\mathbb{Z}_2$}{Z2} number calculation}
	
	There are several methods for calculating the $\mathbb{Z}_2$ number, including Pfaffian method \cite{PhysRevLett.95.226801, PhysRevLett.95.146802, PhysRevB.76.045302}, Wannier function center (WFC) method \cite{PhysRevB.83.035108, PhysRevB.83.235401, PhysRevB.95.075146, PhysRevB.84.075119}, and Berry connection and Berry curvature method \cite{PhysRevB.74.195312, doi:10.1143/JPSJ.76.053702, FENG20121849, PhysRevLett.105.096404}. For numerical calculations, the Fukui-Hatsugai method \cite{doi:10.1143/JPSJ.76.053702} is a direct method, but this method requires gauge fix on the boundary, which means it needs to consider translational symmetry, time reversal symmetry, and Kramar degenerates. The WFC method is an indirect method that does not require any gauges, and the method from Ref.\cite{PhysRevB.84.075119} is used in TopoTB software package. In addition to these, the Shiozaki method \cite{2023arXiv230505615S}, which does not require any gauges, is also adopted by TopoTB software package.
	
	\subsubsection{Wilson loop method}
	
	Here is the main formula for calculating $\mathbb{Z}_2$ number through the U(2N) non-Abelian Berry connection \cite{PhysRevB.84.075119}. The non-Abelian Berry's connection is represented as
	\begin{align}
		F_{i,i+1}^{nm}&=\sum_\alpha c_{n\alpha}^{*}(k_{x,i},k_y) c_{m\alpha}(k_{x,i+1},k_y) \\
		&=\langle n,k_{x,i},k_y | m,k_{x,i+1},k_y \rangle
	\end{align}
	where $n(m)=1,2,\cdots,2N$ represents the sum of 2N occupied states, $k_{x,i}=2\pi i/N_x a_x$ are the discrete points in the $k_x$ direction, $| m,k_{x,i+1},k_y \rangle$ is the periodic part of the intrinsic Bloch function. Connect and multiply $F_{i,i+1}$ end-to-end, and define the $W(k_y)$ matrix as
	\begin{align} \label{eq17}
		W(k_y)=F_{0,1}F_{1,2}F_{2,3} \cdots F_{N_x-2,N_x-1}F_{N_x-1,0}
	\end{align}
	The $W(k_y)$ matrix has 2N eigenvalues
	\begin{align}
		\lambda_n(k_y)=|\lambda_n| e^{i \phi_n (k_y)}
	\end{align}
	where $|\lambda_n|=1$, the phase $\phi_n (k_y)$ of the eigenvalues is the center of the Wanier function for the occupied state of the equivalent one-dimensional system. Ref.\cite{PhysRevB.84.075119} proves that $\phi_n (k_y)$ and $\mathbb{Z}_2$ number are equivalent. By keeping $\phi_n (k_y=0)$ and $\phi_n (k_y=\pi)$ within the $[0,2\pi]$ range, it can be obtained that
	\begin{align}
		\int_0^{\pi} dk_y \partial_{k_y} = \phi_n (\pi) - \phi_n (0) + 2\pi M_n
	\end{align}
	where $M_n$ represents the winding number of $\phi_n (k_y)$, and $\mathbb{Z}_2$ number can be expressed as
	\begin{align}
		\mathbb{Z}_2 = \sum_{n=1}^{2N} M_n\ \operatorname{mod} \ 2
	\end{align} 
	
	\subsubsection{Shiozaki method}
	
	Calculating the $\mathbb{Z}_2$ number requires the TB model to have time reversal symmetry. We define the time reversal operator as
	\begin{align}\label{eq29}
	 	T=U_T K
	\end{align}
    where $U_T$ is a $2N \times 2N$ unitary matrix that satisfies $(U_T)^{\operatorname{tr}}=-U_T$, and $K$ represents complex conjugation. Time reversal symmetry means that
    \begin{align}\label{eq30}
     	H(-\mathbf{k})=T H(\mathbf{k}) T^{-1}
    \end{align}
 	In the first Brillouin zone of two dimensions, there are 4 time reversal invariant points (8 for three dimensions) that satisfy $-\Gamma_i=\Gamma_i+n_i \mathbf{G}$, where $\mathbf{G}$ is a reciprocal lattice vector and $n_i=0$ or $1$ \cite{PhysRevB.76.045302, PhysRevLett.98.106803, shun2018topological}. At these points, $\Gamma_i=n_i \mathbf{G}/2$, the Hamiltonian satisfies
 	\begin{align}
 		H(\Gamma_i)=T H(\Gamma_i) T^{-1}
 	\end{align}
 	Due to the time reversal symmetry, we only need to perform numerical calculations in half of the first Brillouin zone. We always calculate the $\mathbb{Z}_2$ number on a two-dimensional plane, so we only need 4 time reversal invariant points $\Gamma_i=(0,0), (\pi,0), (0,\pi), (\pi,\pi)$. We introduce the unitary matrix
 	\begin{align}
 		w_{mn}(\mathbf{k})=\langle u_m(-\mathbf{k}) | T | u_n(\mathbf{k}) \rangle
 	\end{align}
	where $n(m)=1,2,\cdots,2N$ represents the sum of 2N occupied states. At 4 time reversal invariant points $\Gamma_i=(0,0), (\pi,0), (0,\pi), (\pi,\pi)$, $w$ is antisymmetric. An antisymmetric matrix can calculate Pfaffian, which can be calculated using \lstinline|ResourceFunction["Pfaffian"][m]| \cite{wimmer2012algorithm} in Mathematica. Define time-reversal polarization on $k_y=0$ and $k_y=\pi$ lines as
	\begin{align}\label{eq33}
		P_{T, x}\left(\Gamma_{y}\right)=\frac{1}{2 \pi} \operatorname{Arg} \left[ \left[\prod_{k_{x}=0}^{\pi-\delta k_{x}} \operatorname{Det}\left[\langle u\left(k_{x}+\delta k_{x}, \Gamma_{y}\right) | u\left(k_{x}, \Gamma_{y}\right) \rangle \right] \right] \frac{\operatorname{Pf}\left[w\left(0, \Gamma_{y}\right)\right]}{\operatorname{Pf}\left[w\left(\pi, \Gamma_{y}\right)\right]}\right]
	\end{align}
	where $\Gamma_{y}=0, \pi$. We define Berry flux as $\tilde{F}_l$, as given by Equation(\ref{eq20}) mentioned earlier. By integrating in half of the first Brillouin zone, the total Berry flux is
	\begin{align}\label{eq34}
		\tilde{F}=\sum_l \tilde{F}_l
	\end{align}
	where $l \in [0, 2 \pi]\times [0,\pi]$. From Equation(\ref{eq33}) and Equation(\ref{eq34}), the $\mathbb{Z}_2$ number is given by
	\begin{align}
		\mathbb{Z}_2 = \frac{1}{2\pi}\tilde{F}-2P_{T, x}(0)+2P_{T, x}(\pi) \ \operatorname{mod}\ 2
	\end{align}
	Calculating the $\mathbb{Z}_2$ number using the Shiozaki method requires us to use the ``lattice gauge", otherwise the time reversal invariant point is incorrect. Meanwhile, we also need to determine the time reversal operator and occupied states. Once these are determined, the efficiency of using this method is high.
	
	\subsubsection{3D Brillouin zone}
	
	In a three-dimensional system, there are 8 time reversal invariant points and 6 time reversal invariant planes, labeled as $(x_0,x_\pi,y_0,y_\pi,z_0,z_\pi)$. The $\mathbb{Z}_2$ number of these 6 planes satisfies
	\begin{align}
		x_0+x_\pi=y_0+y_\pi=z_0+z_\pi \ (\operatorname{mod}\ 2)
	\end{align}
	Therefore, only 4 invariants become independent. We define the $\mathbb{Z}_2$ number of the 3D system as
	\begin{align}
		\mathbb{Z}_2=(\nu_0,\nu_1,\nu_2,\nu_3)=(x_0+x_\pi,x_\pi,y_\pi,z_\pi)
	\end{align}
	Specifically, when $v_0=1$, the system is called a strong topological insulator \cite{PhysRevB.76.045302, PhysRevLett.98.106803, PhysRevB.75.121306, shun2018topological, openmx}.
	
	\section{Capabilities of TopoTB}
	\label{Capabilities}
	
	\begin{figure}
		\centering
		\includegraphics[scale=0.85]{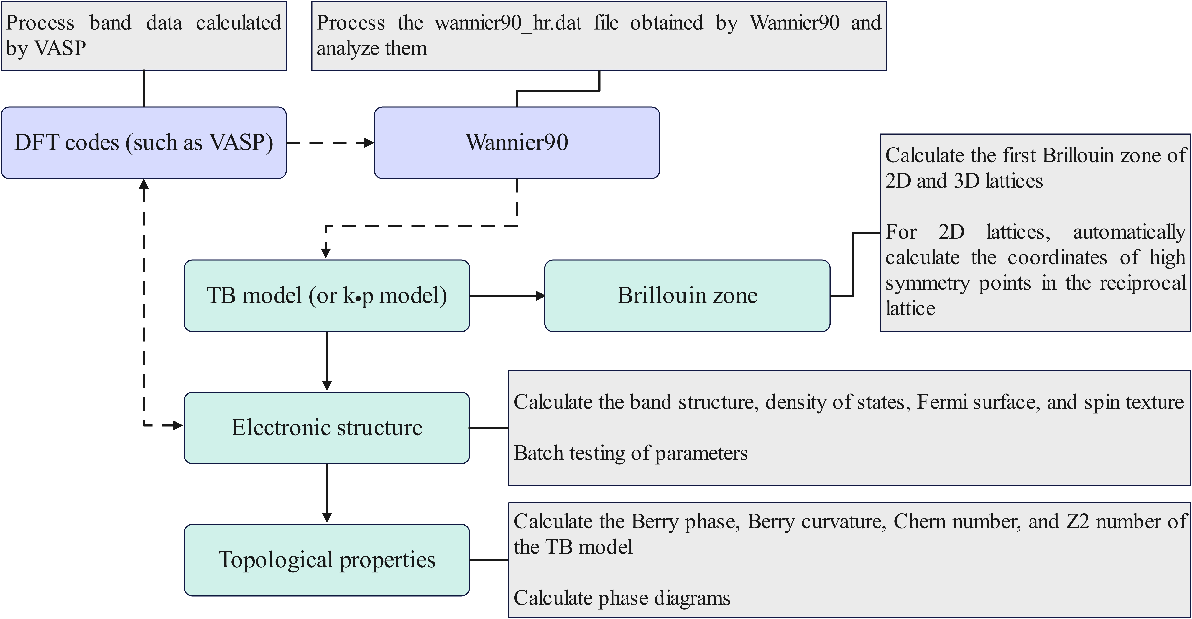}
		\caption{
		The workflow and main capabilities of the TopoTB software package.
		}
		\label{fig2}
	\end{figure}
	
	The main capabilities of the TopoTB software package are electronic structure and topological properties calculation. In addition, it can also provide a schematic diagram of the Brillouin zone. Specifically, for two-dimensional lattices, it can provide the coordinates of the high symmetry point in the first Brillouin zone, which is particularly useful for irregular two-dimensional lattices. The high symmetry point coordinates it provides can assist WannierTools \cite{WU2018405} in calculating surface states, and so on. The TopoTB software package requires the model Hamiltonian, which can be obtained by extracting the results (\lstinline{wannier90_hr.dat} file) calculated by Wannier90 \cite{MOSTOFI2008685} using our software package, or by using software such as MagneticTB \cite{ZHANG2022108153} and MagneticKP \cite{zhang2023magnetickp} software packages. Of course, the TB model can also be built by oneself, and our software package can also accept such Hamiltonians as inputs. The band structure of the TB model may need to be compared with the band structure obtained from DFT codes such as VASP \cite{PhysRevB.54.11169, KRESSE199615}. At present, our software package can process the band data of VASP or the band data obtained from VASPKIT \cite{WANG2021108033} post-processing. In addition to the conventional band structure, the spin polarized band structure under SOC can also be obtained using our software package. Specifically, our software package also includes numerous post-processing codes, allowing for interactive operation of the TB model to determine parameters, compute phase diagrams, and so forth. The main capabilities and possible workflow of the TopoTB software package are shown in Fig.\ref{fig2}.
	
	\section{Installation and usage}
	\label{Installation}
	\subsection{Installation}
	
	To use the TopoTB software package, you must specify the full path to load the software package via \lstinline|Get| or \lstinline|Needs|. Alternatively, if the TopoTB software package is located in a directory included in \lstinline|$Path|, you can directly call it using \lstinline|Get| or \lstinline|Needs|. For convenience, either of the following two lines of code can quickly open a directory contained in \lstinline|$Path|, place the TopoTB software package in it, and call it directly using \lstinline|Get| or \lstinline|Needs|.
	\begin{lstlisting}
(*Select one of the paths provided after running $Path*)
$Path
		
(*Run any of the following two lines of code*)
SystemOpen[FileNameJoin[{$UserBaseDirectory,"Applications"}]]
SystemOpen[FileNameJoin[{$BaseDirectory,"Applications"}]]
	\end{lstlisting}
	Utilize the \lstinline|FindFile| function to locate the installation path of the TopoTB software package and verify whether it is installed in the directory specified in \texttt{\textdollar Path}.
	\begin{lstlisting}
FindFile["TopoTB`"]
	\end{lstlisting}
	
	\subsection{Usage}
	
	\begin{figure}
		\centering
		\includegraphics[scale=0.85]{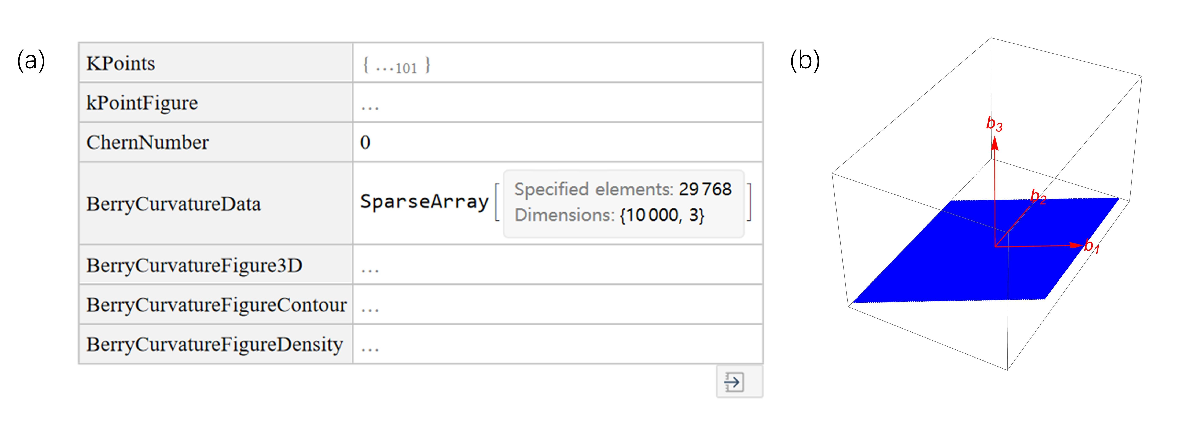}
		\caption{
			(a) The output of the \lstinline|ChernNumberCalc| function, multiple results are displayed in \lstinline|Dataset|. (b) Corresponding to the result in the 2nd row of (a).
		}
		\label{fig3}
	\end{figure}
	Load the TopoTB software package using either \lstinline|Get| or \lstinline|Needs|:
	\begin{lstlisting}
(*Run any of the following two lines of code*)
<<TopoTB`
Needs["TopoTB`"]
	\end{lstlisting}
	Get help by running the \lstinline|?+function|. For example, if you input \lstinline|?ChernNumberCalc|, you will receive help related to the \lstinline|ChernNumberCalc| function, which explains the inputs and outputs of the function. Please refer to \ref{C4} for details.
	
	Using the Hamiltonian of Equation(\ref{eq5}), input the code in the following format to calculate the Chern number:
	\begin{lstlisting}
(*Hamiltonian*)
k={kx,ky};
\[Delta]1=1/2 {Sqrt[3],-1};\[Delta]2={0,1};\[Delta]3=1/2 {-Sqrt[3],-1};
H={{M,0},{0,-M}}+
	{{0,t*Exp[I*k.\[Delta]1]+t*Exp[I*k.\[Delta]2]+t*Exp[I*k.\[Delta]3]},
	{t*Exp[-I*k.\[Delta]1]+t*Exp[-I*k.\[Delta]2]+t*Exp[-I*k.\[Delta]3],0}};
(*Defined function*)
h[{kx_,ky_,kz_},t_:1,M_:1]=H;
(*Direct lattice*)
lat={{Sqrt[3],0,0},1/2 {-Sqrt[3],3,0},{0,0,1}};
(*Band index*)
bnd=Table[i,{i,1,1}];
(*Chern number*)
ds=ChernNumberCalc[h,lat,bnd,{100,100}]
	\end{lstlisting}
	The results obtained are displayed in \lstinline|Dataset| as shown in Fig. \ref{fig3}(a), named \lstinline|ds|. To extract specific results from this dataset, we can utilize \lstinline|Normal[ds[row]]|. For instance, to retrieve the range of the calculated point for the 2nd row, we input \lstinline|Normal[ds[2]]|, as demonstrated in Fig. \ref{fig3}(b). Similarly, for the Berry curvature depicted in Fig. \ref{fig1}(d), the result can be obtained by entering \lstinline|Normal[ds[6]]|. By substituting Equation (\ref{eq5}) with Equation (\ref{eq3}) and using analogous input, we can generate Fig. \ref{fig1}(c).
	
	In summary, in the TopoTB software package, it is essential to provide the Hamiltonian, lattice vectors, and other necessary inputs, and then output significant data, internally defined plots, etc., in the \lstinline|Dataset| format. This format allows for the utilization of the data for replotting purposes.
	
	\section{Examples}
	\label{Examples}
	
	In this section, we demonstrate the capabilities of the TopoTB software package through a few examples. For more examples, please refer to the public code repository: \href{https://github.com/xlhuang-phy/TopoTB/tree/main/examples}{https://github.com/xlhuang-phy/TopoTB/tree/main/examples}.
	
	\subsection{Band structure and density of states}
	
	The Kane-Mele model \cite{PhysRevLett.95.226801, PhysRevLett.95.146802} is a graphene model for describing the quantum spin Hall effect, which has the time reversal symmetry,
	\begin{align}\label{eq38}
		H=t \sum_{\langle i, j\rangle} c_{i}^{\dagger} c_{j}+i \lambda_{{SO}} \sum_{\langle \langle i, j\rangle\rangle} v_{{ij}} c_{i}^{\dagger} s^{z} c_{j}+i \lambda_{R} \sum_{ \langle i, j\rangle} c_{i}^{\dagger}\left(\mathbf{s} \times \hat{\mathbf{d}}_{ij}\right)_{z} c_{j}+\lambda_{v} \sum_{i} \xi_{i} c_{i}^{\dagger} c_{i}
	\end{align}
	where $c_i^{\dagger}=(c_{i,\uparrow}^{\dagger},c_{i,\downarrow}^{\dagger})$. The first term is the nearest neighbor hopping term, which is equivalent to considering SOC in Equation(\ref{eq3}) or Equation(\ref{eq5}). The second item considers the mirror symmetric SOC in the second nearest neighbor, where $v_{{ij}}=\frac{2}{\sqrt{3}}(\hat{\mathbf{d}}_i \times \hat{\mathbf{d}}_j)_z=\pm 1$, where $\hat{\mathbf{d}}_i$ and $\hat{\mathbf{d}}_j$ are two unit vectors along the two bonds, and an electron hopping from $j$ to $i$. $s^z$ is the Pauli matrix that describes the $z$-component spin of an electron. The third item is the Rashba SOC term that breaks the inversion symmetry, $\mathbf{s}$ is the Pauli matrix with $\mathbf{s}=(s^x,s^y,s^z)$. The fourth term is the staggered sublattice potential, where $\xi_{i}=\pm 1$.
	
	To be consistent with Ref.\cite{PhysRevLett.95.146802}, we consider ``lattice gauge". Perform a Fourier transform on Equation(\ref{eq38}) using Equation(\ref{eq2}), taking the lattice vectors $\mathbf{a}_1=(\frac{1}{2},\frac{\sqrt{3}}{2})$ and $\mathbf{a}_2=(-\frac{1}{2},\frac{\sqrt{3}}{2})$, and write the Hamiltonian as
	\begin{align}
		H(\mathbf{k})=\sum_{a=1}^5 d_a(\mathbf{k})\Gamma^a + \sum_{a < b=1}^5 d_{ab}(\mathbf{k})\Gamma^{ab}
	\end{align}
	Here, 5 Dirac matrices and their 10 commutators are as follows:
	\begin{gather}
		\Gamma^{(1,2,3,4,5)}=(\sigma^x \otimes I, \sigma^z \otimes I, \sigma^y \otimes s^x, \sigma^y \otimes s^y, \sigma^y \otimes s^z) \\
		\Gamma^{ab}=[\Gamma^a,\Gamma^b]/(2i)
	\end{gather}
	where $I$ denotes the $2 \times 2$ identity matrix, the Pauli matrices $\sigma^k$ and $s^k$ represent the sublattice and spin indices. Under the Pauli-Dirac representation, the four-band Hamiltonian is expressed by the Dirac matrices. The coefficient in the Hamiltonian are
	\begin{align}
		d_1 &= t(1+2\cos x \cos y ) \\
		d_2 &=\lambda_v \\
		d_3 &= \lambda_R(1-\cos x \cos y ) \\
		d_4 &= -\sqrt{3}\lambda_R\sin x \sin y \\
		d_{12} &= -2t\cos x \sin y  \\
		d_{15} &= \lambda_{SO}(2\sin 2x - 4\sin x \cos y) \\
		d_{23} &= -\lambda_R\cos x \sin y \\
		d_{24} &= \sqrt{3}\lambda_R\sin x \cos y
	\end{align}
	where $x=k_x/2$ and $y=\sqrt{3}k_y/2$. Next, we will investigate the band structure of the Kane-Mele model using the TopoTB software package.
	
	\begin{figure}
		\centering
		\includegraphics[scale=0.85]{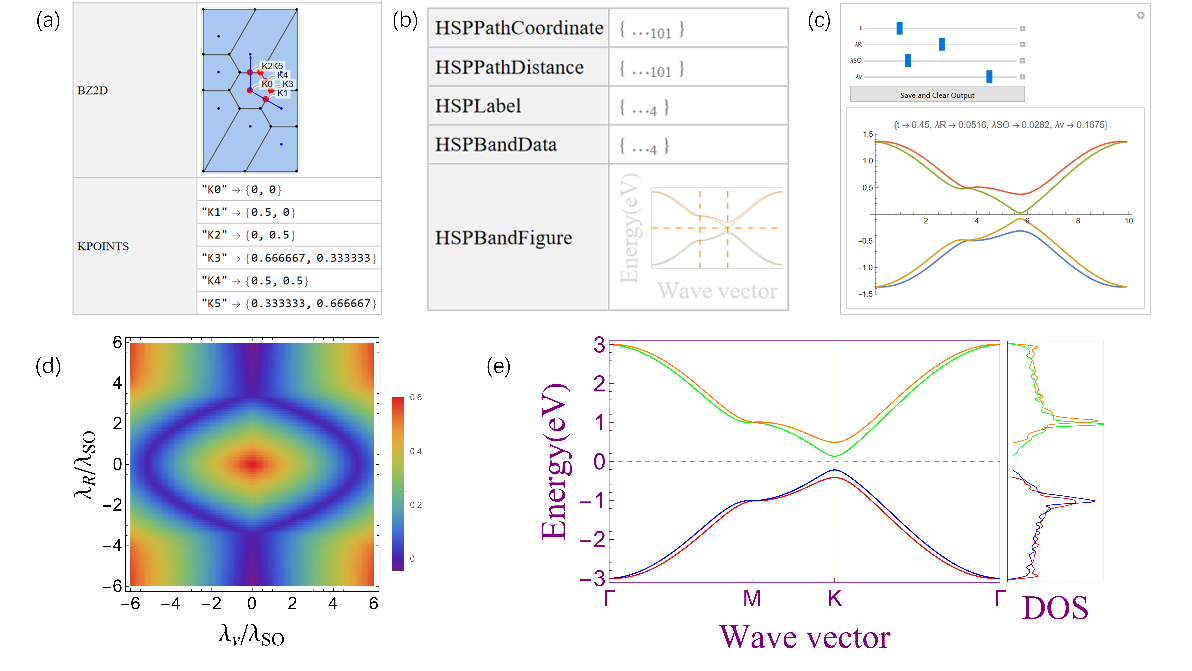}
		\caption{
			(a) Reciprocal lattice and high symmetry point coordinates. (b) Band structure and data along the high symmetry path. (c) Interactive operation to adjust the band structure. (d) Band gap phase diagram. (e) The band structure along a high symmetry path and their density of states.
		}
		\label{fig4}
	\end{figure}
	
	The Kane-Mele model is based on a hexagonal honeycomb lattice. Now, we calculate its reciprocal lattice and high symmetry points using the \lstinline|BrillouinZone2D| function. Since the direct lattice of the Kane-Mele model is in two-dimensional form, we need to transform it into three-dimensional form as input for the \lstinline|BrillouinZone2D| function. The relevant code can be found in \ref{A1}. In the code, we choose the plane determined by $\mathbf{a}_1$ and $\mathbf{a}_2$ as the direct lattice vectors of the Kane-Mele model. The output result is shown in Fig.\ref{fig4}(a). As can be seen, it provides the coordinates of all high symmetry points of the reciprocal lattice, which is very useful for irregular lattices. In addition, calculating the surface state of a three-dimensional lattice requires determining the surface and its corresponding reciprocal lattice through the direct lattice vectors of the three-dimensional lattice, which can be conveniently calculated using this function.
	
	After obtaining the coordinates of the high symmetry points, we can calculate the band structure along the high symmetry path. Define the Hamiltonian of the Kane-Mele model as a function, which is defined in the software package, so it is directly introduced in \ref{A2}. Of course, it can also be manually written out. Then, write down the coordinates of the high symmetry point and the number of points between the two points to calculate directly. The relevant code is shown in \ref{A2}. The output result is shown in Fig.\ref{fig4}(b). We can use the results of the \lstinline|HSPBand| function for other processing, such as batch testing of band structures, interactive testing of band structures, and calculation of band gap phase diagrams. From Fig.\ref{fig4}(b), we can see that the 5th row shows the band structure, and then we can perform parameter batch testing through the cycle. \ref{A3} shows the band structure when the parameter $t$ takes 1, 2, and 3, while keeping the other parameters unchanged, i.e., taking the default values. Run the code and save the result in the current folder.
	
	For interactive testing of band structures, the \lstinline|HSPBand| function is required. At the same time, the 1st and 2nd rows in the output result of the \lstinline|Manipulate| function are required. Fig.\ref{fig4}(c) shows screenshots of interactive operations on the 4 parameters of the Kane-Mele model. The relevant code is provided in \ref{A4}.
	
	When the band gap of the band structure is closed and reopened, a topological phase transition often occurs. For this purpose, we calculated the band gap phase diagram of the Kane-Mele model, as shown in Fig.\ref{fig4}(d). The method is to calculate the difference between the minimum value of the conduction band and the maximum value of the valence band. A positive value indicates an open band gap, while a negative value or zero indicates a closed band gap. The relevant code is provided in \ref{A5}.
	
	In addition to analyzing band structures, studying the density (DOS) \cite{TORIYAMA2022100002} of states can reveal various intriguing properties, including Van Hove singularities \cite{HAN2024319}, among others. For the Kane-Mele model, we calculate its density of states using the \lstinline|DOS2D| function. The density of states needs to be calculated in the first Brillouin zone, and the number of points needs to be sufficient. Through our calculation, we can see a peak in the density of states at $\pm 1$ eV, corresponding to point M, as shown in Fig.\ref{fig4}(e). The relevant code is provided in \ref{A6}.
	
	\subsection{Fermi surface and spin texture}
	
	\begin{figure}
		\centering
		\includegraphics[scale=0.85]{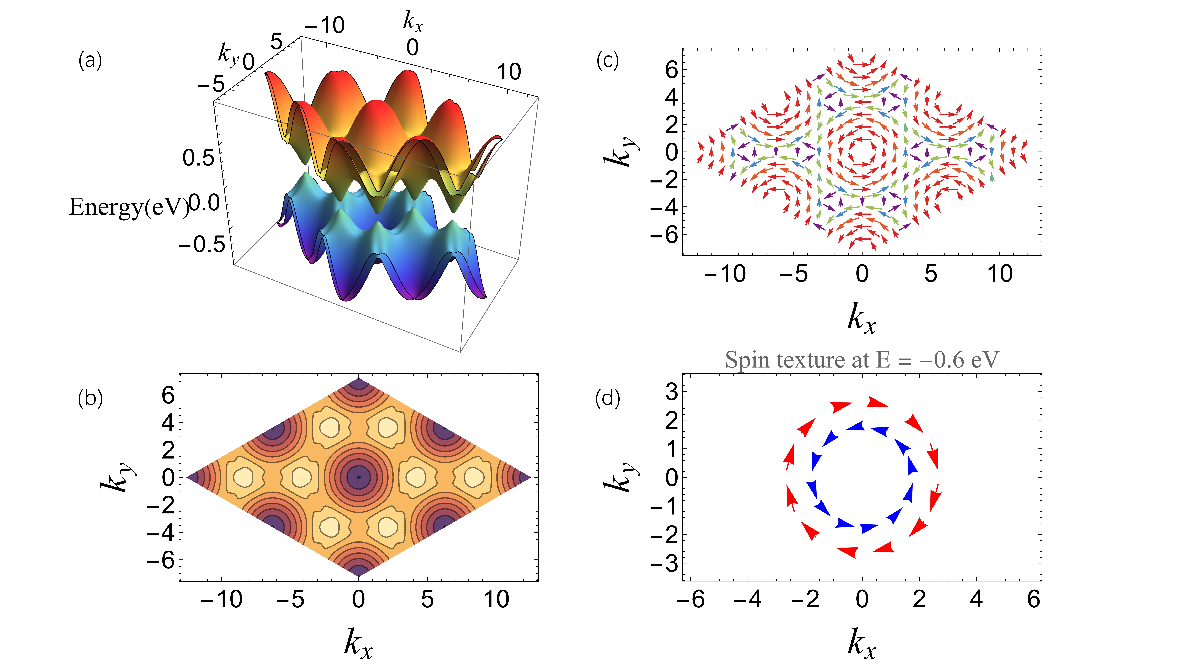}
		\caption{
			(a) Three-dimensional band structure. (b) The Fermi surface of the second band. (c) The spin texture of the second band. (d) The spin texture of the first and second bands at -0.6 eV.
		}
		\label{fig5}
	\end{figure}
	
	Here we will continue to discuss the capabilities of the TopoTB software package in electronic structure calculations. Using the Kane-Mele model, we can calculate its three-dimensional band structure, Fermi surface, and spin texture through the \lstinline|HCalc2D| function. As a demonstration, Fig.\ref{fig5}(a) shows the three-dimensional band structure of the Kane-Mele model, Fig.\ref{fig5}(b) and (c) show the Fermi surface and spin texture of the 2nd band, respectively. The Fermi surface and spin texture of other bands can be viewed through running examples. The parameter values and code corresponding to this result are shown in \ref{B1}.
	
	The Fermi surface and spin texture provided above are global. In addition, our software package can also calculate the Fermi surface and spin texture at specified energy levels. Through the calculation of the \lstinline|SpinTextureAtEnergy2D| function, Fig.\ref{fig5}(d) shows the spin texture of the first and second bands at -0.6 eV. As can be seen, this is a spin texture with opposite chirality, corresponding to the conventional Rashba effect. The relevant codes are shown in \ref{B2}.
	
	\subsection{Chern number and \texorpdfstring{$\mathbb{Z}_2$}{Z2} number}
	\subsubsection{\texorpdfstring{$\mathbb{Z}_2$}{Z2} number}
	
	\begin{figure}
		\centering
		\includegraphics[scale=0.85]{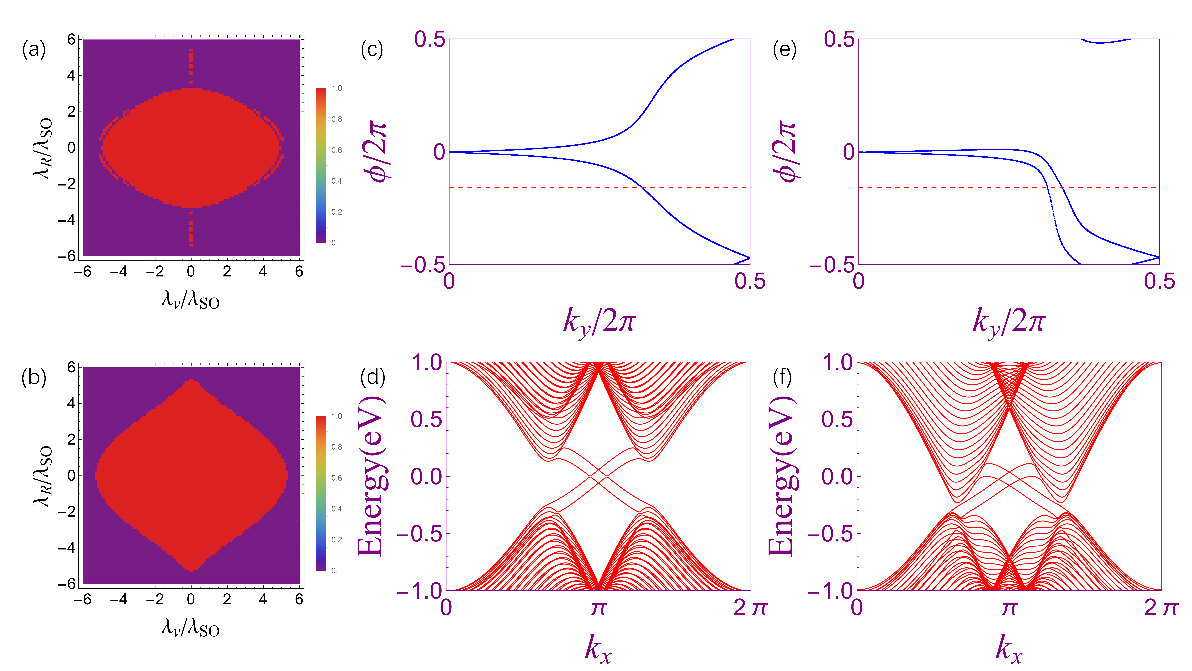}
		\caption{
			(a) and (b) are $\mathbb{Z}_2$ number phase diagrams calculated using the Wilson loop method and Shiozaki method, respectively. The phase diagrams are functions of $\lambda_{v}$ and $\lambda_{R}$. (c) and (e) are the evolution lines of the Wannier centers, corresponding to the (1,1) and (1,4) coordinate points in the phase diagram, respectively. (d) and (f) are the band structures of one-dimensional zigzag edges, corresponding to the (1,1) and (1,4) coordinate points in the phase diagram, respectively. In all cases, take $t=1 $ and $\lambda_{{SO}}=0.06 $.
		}
		\label{fig6}
	\end{figure}
	
	In the TopoTB software package, we implemented the Wilson loop method \cite{PhysRevB.84.075119} and Shiozaki method \cite{2023arXiv230505615S} to calculate the $\mathbb{Z}_2$ number, which correspond to indirect and direct calculations, respectively. The $\mathbb{Z}_2$ number phase diagrams of the Kane-Mele model were calculated using these two methods, i.e, \lstinline|Z2NumberCalcByWilsonLoop| and \lstinline|Z2NumberCalc|, corresponding to Fig.\ref{fig6}(a) and Fig.\ref{fig6}(b), respectively. In Fig.\ref{fig4}(d), we calculated the band gap phase diagram of the Kane-Mele model and combined it with the $\mathbb{Z}_2$ number phase diagram here. It can be seen that the two phase diagrams are different at the boundary. The relevant code for calculating the $\mathbb{Z}_2$ number phase diagram using the Wilson loop method is provided in \ref{C1}. The relevant code for calculating the $\mathbb{Z}_2$ number phase diagrams using the Shiozaki method is provided in \ref{C2}.
	
	To implement the Shiozaki method, we need to provide the time reversal operator. For the Kane-Mele model, the corresponding the time reversal operator is $T=\sigma_0 \otimes i \sigma_{y} K$, where $\sigma_i$ is the Pauli matrix, so the $U_T$ in Equation(\ref{eq29}) is
	\begin{lstlisting}
(*Time reversal operator*)
UT=KroneckerProduct[PauliMatrix[0],I*PauliMatrix[2]]
	\end{lstlisting}
	The Hamiltonian is $H$, and the time reversal symmetry is verified through Equation(\ref{eq30}). The corresponding code is
	\begin{lstlisting}
ComplexExpand[ConjugateTranspose[UT].Conjugate[H/.{kx->-kx,ky->-ky}].UT]
	\end{lstlisting}
	Referring to Ref.\cite{PhysRevLett.95.146802}, take $t=1 $ and $\lambda_{{SO}}=0.06$, and the $\mathbb{Z}_2$ number phase diagram is a function of $\lambda_{v}$ and $\lambda_{R}$. Taking coordinates (1,1) in the phase diagram, i.e., $\lambda_{v}=0.06$ and $\lambda_{R}=0.06$, both methods yield topologically nontrivial results. We see that the reference line in Fig.\ref{fig6}(c) intersects with the evolution line of the Wannier centers, indicating that the $\mathbb{Z}_2$ number is 1. Fig.\ref{fig6}(d) shows the edge states corresponding to Fig.\ref{fig6}(c), which is consistent with its results. However, taking coordinates (1,4) in the phase diagram, i.e., $\lambda_{v}=0.06$ and $\lambda_{R}=0.24$, the results obtained by the two methods are different. For the Wilson loop method, the evolution line of the Wannier centers obtained is shown in Fig.\ref{fig6}(e), with two intersection points between the reference line and it, corresponding to topologically trivial. The edge states obtained from this set of parameters are shown in Fig.\ref{fig6}(f), and it can be seen that there are edge states. Therefore, near the band gap, the Wilson loop method may encounter issues, while the Shiozaki method is a better approach.
	
	\subsubsection{Chern number}
	
	\begin{figure}
		\centering
		\includegraphics[scale=0.85]{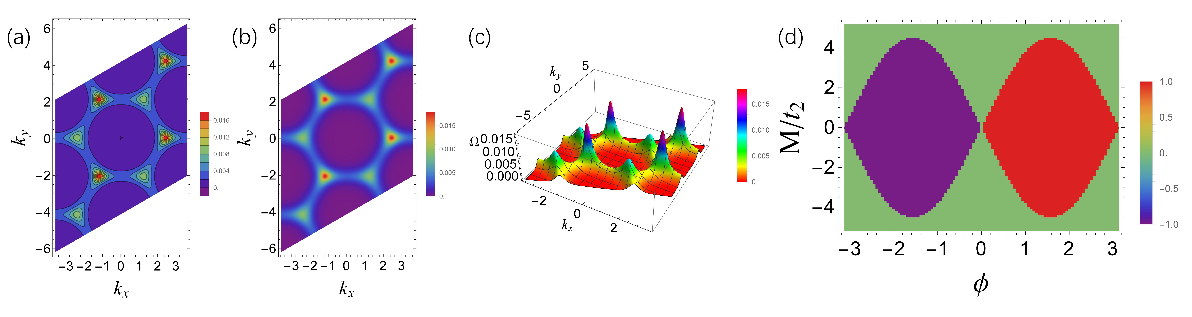}
		\caption{
			(a) and (b) show two styles of two-dimensional Berry curvature plots. (c) shows a three-dimensional Berry curvature plot. (d) is the Chern number phase diagram of the Haldane model, and the phase diagram is a function of $\phi$ and $M$, where $t_1=1$ and $t_2=1$.
		}
		\label{fig7}
	\end{figure}
	
	The Haldane model \cite{PhysRevLett.61.2015} is a graphene model that does not consider SOC and does not have a Landau level, but exhibits integer quantum Hall effect. When the parameter $\lambda_{R}=0$ in the Kane-Mele model, it can be seen as the superposition of two sets of the Haldane model. The Hamiltonian of the Haldane model is written as
	\begin{align}
		H=\epsilon(\mathbf{k})+\mathbf{d}(\mathbf{k}) \cdot \boldsymbol{\sigma}
	\end{align}
	where
	\begin{align}
		\epsilon(\mathbf{k})&=2 t_{2} \cos(\phi) \sum_{i=1,2,3} \cos \left(\mathbf{k} \cdot \mathbf{b}_{i}\right) \\
		d_{x}(\mathbf{k})&=t_{1} \sum_{i=1,2,3} \cos \left(\mathbf{k} \cdot \mathbf{a}_{i}\right) \\
		d_{y}(\mathbf{k})&=t_{1} \sum_{i=1,2,3} \sin \left(\mathbf{k} \cdot \mathbf{a}_{i}\right) \\
		d_{z}(\mathbf{k})&=M-2 t_{2} \sin(\phi) \left(\sum_{i=1,2,3} \sin \left(\mathbf{k} \cdot \mathbf{b}_{i}\right)\right)
	\end{align}
	$(\mathbf{a}_1,\mathbf{a}_2,\mathbf{a}_3)$ is a vector from a site to its nearest neighboring site, and
	\begin{align}
		\mathbf{b}_1=\mathbf{a}_2-\mathbf{a}_3 \qquad \mathbf{b}_2=\mathbf{a}_3-\mathbf{a}_1 \qquad \mathbf{b}_3=\mathbf{a}_1-\mathbf{a}_2
	\end{align}
	$\boldsymbol{\sigma}=(\sigma_x,\sigma_y,\sigma_z)$ is the Pauli matrix. Next, we will calculate the Chern number of the Haldane model and provide a phase diagram about the Chern number.	
	
	By defining the Hamiltonian of the Haldane model as a function and inputting the direct lattice vector and occupied states, we will obtain an output similar to Fig.\ref{fig2}(a). Taking $t_1=1,t_2=0.2,M=0.1,\phi=0.2\pi$, the calculated Chern number is 1, through the \lstinline|ChernNumberCalc| function, and the corresponding Berry curvature is shown in Fig.\ref{fig7}(a), (b), and (c). In order to understand the topological phase transitions in the Haldane model, we calculated the topological phase transitions caused by $\phi$ and $M$ changes at $t_1=1$ and $t_2=1$ using the TopoTB software package. The phase diagram is shown in Fig.\ref{fig7}(d), and the relevant code is provided in \ref{C3}.
	
	\subsection{\texorpdfstring{$\mathbf{k} \cdot \mathbf{p}$}{k dot p} model}
	
	\begin{figure}
		\centering
		\includegraphics[scale=0.85]{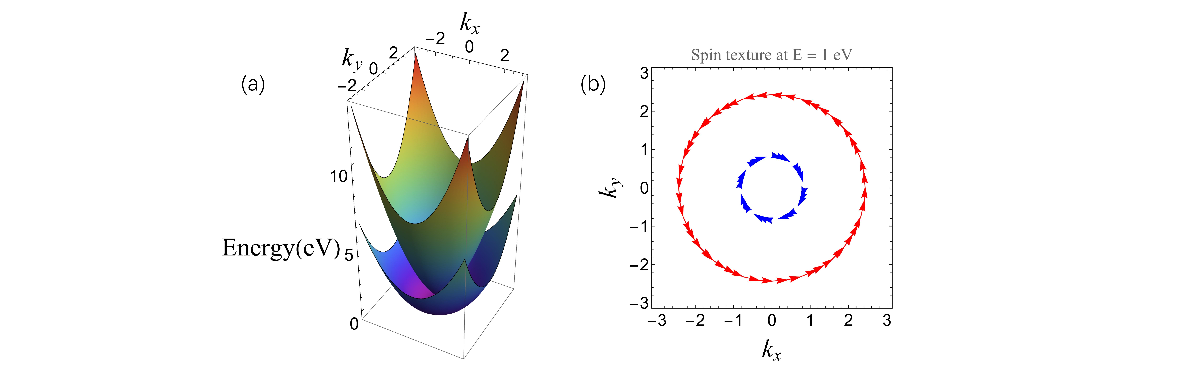}
		\caption{
			(a) The three-dimensional band structure of the $\mathbf{k} \cdot \mathbf{p}$ model. (b) The spin texture of the $\mathbf{k} \cdot \mathbf{p}$ model at 6 eV.
		}
		\label{fig8}
	\end{figure}
	
	The TopoTB software package can not only calculate the TB model, but also the $\mathbf{k} \cdot \mathbf{p}$ model. For the $\mathbf{k} \cdot \mathbf{p}$ model, there is no concept of Brillouin zone. We can obtain the reciprocal lattice vector by setting the direct lattice vector, and thus calculate the electronic structure of the $\mathbf{k} \cdot \mathbf{p}$ model in the momentum space. As an example, we consider a simple $\mathbf{k} \cdot \mathbf{p}$ model \cite{bihlmayer2022rashba} where the Hamiltonian is written as
	\begin{align}
		H(\mathbf{k})=\begin{pmatrix}
			\frac{k_x^2+k_y^2}{2m} & i \alpha (k_x-i k_y) \\
			-i \alpha (k_x+i k_y) & \frac{k_x^2+k_y^2}{2m}
		\end{pmatrix}
	\end{align}
	where $m$ is the mass of the electron, and $\alpha$ is the coupling strength of Rashba SOC. Taking $m=1$ and $\alpha=0.8$, we calculate their three-dimensional band structure and spin texture, corresponding to Fig.\ref{fig8}(a) and (b), respectively. The relevant codes are provided in \ref{D1}.
	
	\subsection{Post-processing capabilities}
	
	Now, we demonstrate the post-processing capabilities in the TopoTB package. LaH$_2$ monolayer is an ideal ferrovalley direct semiconductor with room-temperature ferromagnetic stability \cite{Shi_2022}. Without considering SOC, its band structure exhibits spin polarization. When considering SOC, its band structure exhibits valley polarization, and by performing $S_z$ spin projection on the band structure, it can be seen that the conduction and valence bands are fully spin polarized. Next, we will use the TopoTB software package to calculate the band structure and spin projection considering SOC.
	
	\begin{figure}
		\centering
		\includegraphics[scale=0.85]{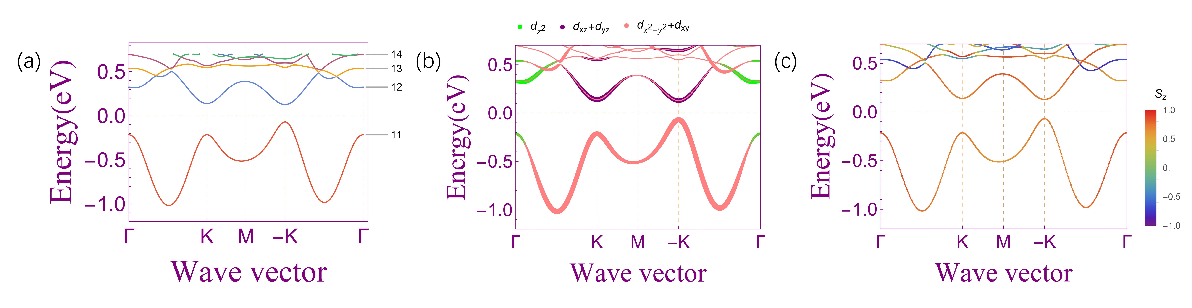}
		\caption{
			(a) The band index band structure of LaH$_2$ monolayer considering SOC. (b) The orbital-resolved band structure of LaH$_2$ monolayer considering SOC. (c) The band structure of LaH$_2$ monolayer with spin $S_z$ projection under considering SOC.
		}
		\label{fig9}
	\end{figure}
	
	\begin{figure}
		\centering
		\includegraphics[scale=0.75]{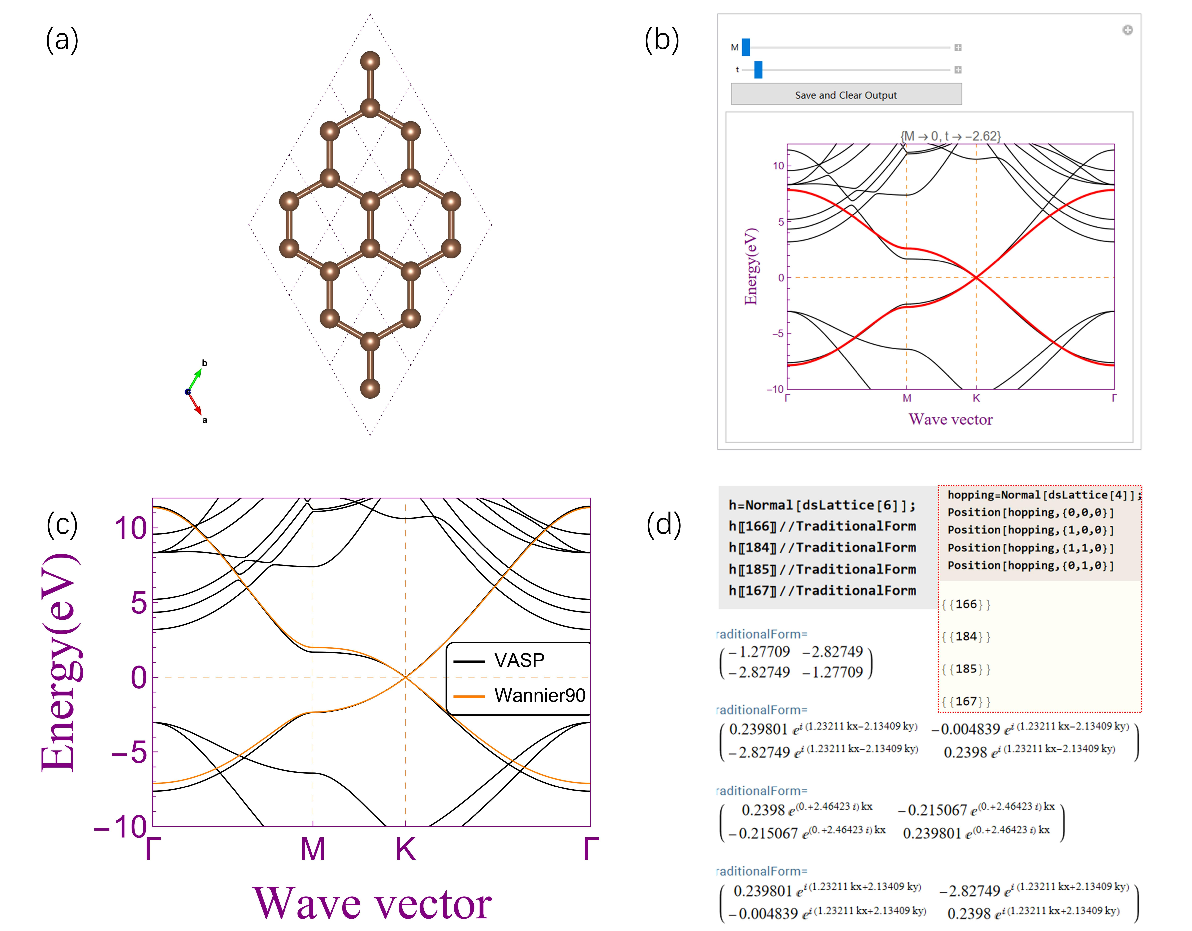}
		\caption{
			(a) Top view of the crystal structure of graphene. (b) Interactive operation of the band structure of the TB model, where the band structure calculated by VASP serves as the background. (c) Comparison of band structures calculated by VASP and Wannier90. (d) Analysis of \lstinline|wannier90_hr.dat| file files calculated by Wannier90
		}
		\label{fig10}
	\end{figure}
	
	Firstly, we perform first principles calculations. Our first principles calculations were based on the density functional theory (DFT), using the projector augmented wave method as implemented in VASP \cite{PhysRevB.49.14251,PhysRevB.54.11169,PhysRevB.50.17953}. The generalized gradient approximation with the Perdew-Burke-Ernzerhof (PBE) \cite{PhysRevLett.77.3865} realization was adopted for the exchange-correlation functional. The cutoff energy was set as 550 eV. The energy and force convergence  criteria were set to be $10^{-8}$ eV and -0.002 eV/\r{A}, respectively. The sampling of the Brillouin zone during the self-consistent process is the grid of $9\times9\times1$.
	
	Secondly, we use the TopoTB software package to process the calculated results. Import \lstinline|POSCAR|, \lstinline|KPOINTS|, and \lstinline|PROCAR| files, input Fermi energy levels, and calculate them using the \lstinline|VASPPBand| function. After the calculation is completed, a \lstinline|.dat| file will be output, and functions existing in the TopoTB software package can also be used to plot. In Fig.\ref{fig9}(a), there is a band structure with band index, which allows you to easily select the band you need. For example, the Fig.\ref{fig9}(a) can intuitively show that the conduction band and valence band correspond to the 11th and 12th bands, respectively. Analyzing the orbital-resolved band structure is crucial. As shown in Fig.\ref{fig9}(b), we can see that the main contribution of the valence band comes from the $d_{x^2-y^2}$ and $d_{xy}$ orbitals. The band structure of LaH$_2$ is fully spin polarized, meaning that near the Fermi surface, the conduction and valence bands are contributed by a single spin. For this, we demonstrate the spin $S_z$ projection of the bands considering SOC, as shown in Fig.\ref{fig9}(c). It can be seen that the band structure of this structure is indeed fully spin polarized. The relevant code is provided in \ref{E1}.
	
	The TopoTB software package can perform interactive operations on the band structure of the TB model, where the band structure calculated by VASP serves as the background. For example, Fig.\ref{fig10}(b) shows the band structure of graphene, and the TB model is similar to Equation(\ref{eq3}), where the lattice constant is taken from the \lstinline|POSCAR| file in VASP, as shown in Fig.\ref{fig10}(a). In this case, the Hamiltonian is expressed as
	\begin{align} \label{vasp}
		H(\mathbf{k})=\begin{pmatrix}
			M & h \\
			h^{*}  & -M
		\end{pmatrix}
	\end{align}
	where $h=te^{-i(1.23211k_x-2.13409k_y)} + te^{i(1.23211k_x+2.13409k_y)} + t$. Through interactive operations, we can take values of $M=0$ and $t=-2.62$.
	
	In addition to processing VASP band data, the TopoTB software package interfaces with Wannier90 \cite{MOSTOFI2008685} to calculate \lstinline{wannier90_hr.dat} files for subsequent analysis and calculation. For instance, Fig.\ref{fig10}(c) displays the band structure calculated by both VASP and Wannier90, demonstrating excellent band fitting near the K point of interest. By analyzing the \lstinline{wannier90_hr.dat} file using TopoTB, we can derive the hopping matrix of the on-site term and its nearest neighbor term, yielding values of $M=0$ and $t=-2.82749$, as shown in Fig.\ref{fig10}(d). Equation (\ref{vasp}), with fewer parameters, facilitates straightforward analysis. However, for TB models with numerous parameters, interactive operations are recommended for determining parameter values. Furthermore, we computed Bi$_2$Se$_3$ \cite{zhang2009topological, xia2009observation} using Wannier90 to obtain its \lstinline{wannier90_hr.dat} file, and subsequently utilized TopoTB to compute its $\mathbb{Z}_2$ number, indicating its classification as a strong topological insulator. For further details, please refer to \href{https://github.com/xlhuang-phy/TopoTB/tree/main/examples}{https://github.com/xlhuang-phy/TopoTB/tree/main/examples}.
	
	\section{Conclusion}
	\label{Conclusion}
	
	The TopoTB software package is a user-friendly software with simple input and convenient output and visualization of data. This software is mainly used to calculate the electronic structure and topological properties of the TB model. In addition, it also interfaces with other software such as VASP, Wannier90, WannierTools, etc. Through some examples, we have demonstrated the main capabilities of software and will continuously improve and expand other capabilities. We hope that this software can become an efficient tool for studying the TB model.
	
	\section*{Acknowledgments}
	\label{ Acknowledgments}
	
	XLH would like to express gratitude to Ken Shiozaki for providing guidance on $\mathbb{Z}_2$ number calculation via email. Additionally, appreciation goes to Yongting Shi for testing and providing suggestions regarding the software package. This work was financially
	supported by the National Key R\&D Program of China (Grants No.
	2022YFA1403200), National Natural Science Foundation of
	China (Grants No. 92265104, No. 12022413, No. 11674331), the Basic Research Program of the Chinese
	Academy of Sciences Based on Major Scientific Infrastructures (Grant No. JZHKYPT-2021-08), the CASHIPS Director’s Fund (Grant No. BJPY2023A09), the \textquotedblleft
	Strategic Priority Research Program (B)\textquotedblright\ of the Chinese
	Academy of Sciences (Grant No. XDB33030100), and the Major Basic Program of Natural
	Science Foundation of Shandong Province (Grant No. ZR2021ZD01). A portion of
	this work was supported by the High Magnetic Field Laboratory of Anhui
	Province, China.
	
	\appendix
	\section{The relevant code input for calculating the band structure and density of states}
	
	\subsection{\lstinline|BrillouinZone2D| function}
	\label{A1}
	\begin{lstlisting}
(*For any two-dimensional surface, this function gives its direct lattice and reciprocal lattice, as well as its high symmetry point coordinates*)
{a1,a2,a3}={{1/2,Sqrt[3]/2,0},{-(1/2),Sqrt[3]/2,0},{0,0,3}};
lat={a1,a2,a3};
(*The two vectors determine which surface to calculate. This can assist WannierTools in calculating the high symmetry point coordinates of the surface states of three-dimensional materials.*)
surf001={a1,a2};
BrillouinZone2D[lat,surf001,{-1,1}]
	\end{lstlisting}

	\subsection{\lstinline|HSPBand| function}
	\label{A2}
	\begin{lstlisting}
Clear[h]
(*Kane-Mele model*)
H=KaneMeleModel[kx,ky,t,\[Lambda]v,\[Lambda]R,\[Lambda]SO];
(*The Hamiltonian is defined as a function of the following format, h[{kx,ky,kz}, parameter...]*)
h[{kx_,ky_,kz_},t_:1,\[Lambda]R_:0.05,\[Lambda]SO_:0.06,\[Lambda]v_:0.1]=H;
(*The coordinates of high symmetry points are written in fractional form*)
G0={0,0,0};K0={0.666667,0.333333,0};M0={0.5,0,0};
(*High symmetry point name used for plot*)
klabel={"\[CapitalGamma]","M","K","\[CapitalGamma]"};
(*This means there are 40 points between \[CapitalGamma] and M, 20 points between M and K, and 40 points between K and \[CapitalGamma].*)
ds=HSPBand[h,lat,{{G0,M0,40},{M0,K0,20},{K0,G0,40}},klabel,0]
	\end{lstlisting}
	
	\subsection{Batch testing}
	\label{A3}
	\begin{lstlisting}
(*Post-processing: cyclic parameter testing*)
Clear[h]
(*Save to the current folder*)
SetDirectory[FileNameTake[NotebookFileName[],{1,-2}]];
Do[
	Module[{bandfig},
		h[{kx_,ky_,kz_},t_:1,\[Lambda]R_:0.05,\[Lambda]SO_:0.06,\[Lambda]v_:0.1]=H;
		bandfig=HSPBand[h,lat,{{G0,M0,40},{M0,K0,20},{K0,G0,40}},klabel][5];
		Export["t="<>ToString[t]<>".png",bandfig,ImageResolution->500]],
	{t,1,3}]
	\end{lstlisting}
	
	\subsection{Interactive testing}
	\label{A4}
	\begin{lstlisting}
(*Post-processing: interactive parametric testing*)
(*Note 1: The 1st and 2nd rows of data from the previous calculation are used*)
(*Note 2: After debugging, click "Save and Clear Output" to save the current picture and delete the output*)
Clear[h]
(*Save to the current folder*)
SetDirectory[FileNameTake[NotebookFileName[],{1,-2}]];
(*Set the correct number of bands*)
nband=4;
kpath=Normal[ds[1]];
kdist=Normal[ds[2]];
Manipulate[
	bandFig=Module[{},
	 	h[{kx_,ky_,kz_},t_:1,\[Lambda]R_:0.05,\[Lambda]SO_:0.06,\[Lambda]v_:0.1]=H;
	 	vals=Table[Eigenvalues[h[i,t,\[Lambda]R,\[Lambda]SO,\[Lambda]v]]//Sort,{i,kpath}];
	 	band=Table[Transpose[{kdist,vals[[All,i]]}],{i,1,nband}];
	 	ListLinePlot[band,PlotLabel->{"t"->ToString[t],"\[Lambda]R"->ToString[\[Lambda]R],"\[Lambda]SO"->ToString[\[Lambda]SO],"\[Lambda]v"->ToString[\[Lambda]v]}]],
	{t,0,2},{\[Lambda]R,0,0.1},{\[Lambda]SO,0,0.1},{\[Lambda]v,0,0.2},
	Button["Save and Clear Output",
		(*Save the picture with the current parameters*)
		fileName="t="<>ToString[t]<>";\[Lambda]R="<>ToString[\[Lambda]R]<>";\[Lambda]SO="<>ToString[\[Lambda]SO]<>";\[Lambda]v="<>ToString[\[Lambda]v]<>".png";
		Export[fileName,bandFig];
		NotebookSave[];
		NotebookDelete[EvaluationCell[]]]
	]
	\end{lstlisting}

	\subsection{Band gap phase diagram}
	\label{A5}
	\begin{lstlisting}
(*Band gap phase diagram*)
\[Lambda]vRange=Subdivide[-0.36,0.36,100]//N;
\[Lambda]RRange=Subdivide[-0.36,0.36,100]//N;
(*Calculate the bandgap of the 2nd and 3rd band*)
Clear[h]
gapData=Table[{\[Lambda]v/0.06,\[Lambda]R/0.06,
	Module[{},
		h[{kx_,ky_,kz_},t_:1,\[Lambda]R_:0.05,\[Lambda]SO_:0.06,\[Lambda]v_:0.4]=H;
		ds=HSPBand[h,lat,{{G0,M0,40},{M0,K0,20},{K0,G0,40}},klabel];
		Min[Normal[ds[4]][[3]][[All,2]]]-Max[Normal[ds[4]][[2]][[All,2]]]]},
	{\[Lambda]v,\[Lambda]vRange},{\[Lambda]R,\[Lambda]RRange}];
(*Phase diagram plot*)
ListDensityPlot[Flatten[gapData//Chop,1],PlotRange->Full,PlotLegends->Automatic,InterpolationOrder->0,ColorFunction->"Rainbow",AspectRatio->1]
	\end{lstlisting}
	
	\subsection{\lstinline|DOS2D| function}
	\label{A6}
	\begin{lstlisting}
(*Band*)
Clear[h]
h[{kx_,ky_,kz_},t_:1,\[Lambda]R_:0.05,\[Lambda]SO_:0.06,\[Lambda]v_:0.1]=H;
lat={{1/2,Sqrt[3]/2,0},{-(1/2),Sqrt[3]/2,0},{0,0,3}};
G0={0,0,0};K0={2/3,1/3,0};M0={1/2,0,0};
klabel={"\[CapitalGamma]","M","K","\[CapitalGamma]"};
dsBand=HSPBand[h,lat,{{G0,M0,40},{M0,K0,20},{K0,G0,40}},klabel,0,{1,2,3,4},{Red,Blue,Green,Orange}];
(*DOS*)
Clear[h]
h[{kx_,ky_,kz_},t_:1,\[Lambda]R_:0.05,\[Lambda]SO_:0.06,\[Lambda]v_:0.1]=H;
lat={{1/2,Sqrt[3]/2,0},{-(1/2),Sqrt[3]/2,0},{0,0,3}};
dsDOS1=DOS2D[h,lat,{100,100},0.08,0,1,Red];
dsDOS2=DOS2D[h,lat,{100,100},0.08,0,2,Blue];
dsDOS3=DOS2D[h,lat,{100,100},0.08,0,3,Green];
dsDOS4=DOS2D[h,lat,{100,100},0.08,0,4,Orange];
figDOS=Show[dsDOS1[[3]],dsDOS2[[3]],dsDOS3[[3]],dsDOS4[[3]]];
(*Band and DOS*)
(*Note: Use the PlotRange function to specify that the range of the y-axis must be the same*)
banddosfig=GraphicsRow[{
	Show[dsBand[5],PlotRange->{Full,{-3.1,3.1}},PlotRangePadding->None],
	Show[figDOS,PlotRange->{Full,{-3.1,3.1}},PlotRangePadding->None]
	},1]
	\end{lstlisting}
	
	\section{The relevant code input for calculating the Fermi surface and spin texture}
	
	\subsection{\lstinline|HCalc2D| function}
	\label{B1}
	\begin{lstlisting}
Clear[h]
h[{kx_,ky_,kz_},t_:0.3,\[Lambda]R_:0.075,\[Lambda]SO_:0.0306,\[Lambda]v_:0.123]=H;
lat={{1/2,Sqrt[3]/2,0},{-(1/2),Sqrt[3]/2,0},{0,0,3}};
ds=HCalc2D[h,lat,{1,2,3,4},{50,50},{2,2},2,1,0];
(*Fermi surface*)
Normal[ds[5]][[2]]
(*Spin texture*)
Normal[ds[6]][[2]]
	\end{lstlisting}
	
	\subsection{\lstinline|SpinTextureAtEnergy2D| function}
	\label{B2}
	\begin{lstlisting}
Clear[h]
h[{kx_,ky_,kz_},t_:0.3,\[Lambda]R_:0.075,\[Lambda]SO_:0.0306,\[Lambda]v_:0.123]=H;
lat={{1/2,Sqrt[3]/2,0},{-(1/2),Sqrt[3]/2,0},{0,0,3}};
(*At E=-0.6, the 1st band, red spin texture*)
ds1=SpinTextureAtEnergy2D[h,lat,-0.6,1,Red,1,0.006,1,{40,40}];
(*At E=-0.6, the 2nd band, blue spin texture*)
ds2=SpinTextureAtEnergy2D[h,lat,-0.6,2,Blue,1,0.006,1,{30,30}];
(*The 1st and 2nd bands are the spin texture at E=-0.6*)
Show[ds1[2],ds2[2]]
	\end{lstlisting}
	
	\section{The relevant code input for calculating the Chern number and $\mathbb{Z}_2$ number}
	
	\subsection{$\mathbb{Z}_2$ number phase diagram by Wilson loop method}
	\label{C1}
	\begin{lstlisting}
(*Z2 phase diagram*)
\[Lambda]vRange=Subdivide[-0.36,0.36,100]//N;
\[Lambda]RRange=Subdivide[-0.36,0.36,100]//N;
(*Calculate Z2*)
Clear[h]
z2Data=Table[{\[Lambda]v/0.06,\[Lambda]R/0.06,
	Module[{},h[{kx_,ky_,kz_},t_:1,\[Lambda]R_:0.05,\[Lambda]SO_:0.06,\[Lambda]v_:0.4]=H;Z2NumberCalcByWilsonLoop[h,lat,{1,2},5,-1,10^-1,10^-2,10^-4]][3]},
	{\[Lambda]v,\[Lambda]vRange},{\[Lambda]R,\[Lambda]RRange}];
(*Z2 phase diagram plot*)
ListDensityPlot[Flatten[z2Data//Chop,1],PlotRange->Full,PlotLegends->Automatic,InterpolationOrder->0,ColorFunction->"Rainbow",AspectRatio->1]
	\end{lstlisting}
	
	\subsection{$\mathbb{Z}_2$ number phase diagram by Shiozaki method}
	\label{C2}
	\begin{lstlisting}
(*Z2 phase diagram*)
\[Lambda]vRange=Subdivide[-0.36,0.36,100]//N;
\[Lambda]RRange=Subdivide[-0.36,0.36,100]//N;
(*Calculate Z2*)
Clear[h]
z2Data=Table[{\[Lambda]v/0.06,\[Lambda]R/0.06,
	Module[{},h[{kx_,ky_,kz_},t_:1,\[Lambda]R_:0.05,\[Lambda]SO_:0.06,\[Lambda]v_:0.4]=H;Z2NumberCalc[h,lat,{1,2},UT]][3]},
	{\[Lambda]v,\[Lambda]vRange},{\[Lambda]R,\[Lambda]RRange}];
(*Z2 phase diagram plot*)
ListDensityPlot[Flatten[z2Data//Chop,1],PlotRange->Full,PlotLegends->Automatic,InterpolationOrder->0,ColorFunction->"Rainbow",AspectRatio->1]
	\end{lstlisting}
	
	\subsection{Chern number phase diagram}
	\label{C3}
	\begin{lstlisting}
(*Haldane model*)
HHaldane=HaldaneModel[kx,ky,t1,t2,M,\[Phi]];
h[{kx_,ky_,kz_},t1_:1,t2_:0.2,M_:0.1,\[Phi]_:0.2*Pi]=HHaldane;
lat={{-Sqrt[3],0,0},{Sqrt[3]/2,-(3/2),0},{0,0,1}};
bnd=Table[i,{i,1,1}];
MRange=Subdivide[-3 Sqrt[3],3 Sqrt[3],30]//N;
\[Phi]Range=Subdivide[-Pi,Pi,30]//N;
(*Calculate Chern number*)
Clear[h]
phaseDiagramData=Table[{\[Phi],M,
	Module[{},h[{kx_,ky_,kz_},t1_:1,t2_:1,M_:0.1,\[Phi]_:0.2*Pi]=HHaldane;ChernNumberCalc[h,lat,bnd,{5,5},{1,1}]][3]},
	{\[Phi],\[Phi]Range},{M,MRange}];
(*Chern number phase diagram plot*)
ListDensityPlot[Flatten[phaseDiagramData//Chop,1],PlotRange->Full,PlotLegends->Automatic,InterpolationOrder->0,ColorFunction->"Rainbow",AspectRatio->3/4]
	\end{lstlisting}
	
	\subsection{\lstinline|?ChernNumberCalc|}
	\label{C4}
\begin{lstlisting}
ChernNumberCalc[H_,lattice_,bandIndex_,b1b2points_:{50,50},b1b2ranges_:{2,2}]
	
Input:
H: The format of the Hamiltonian is H[{kx, ky, kz}, parameter...].
lattice: The primitive vectors of Bravais Lattices (Lattice Vectors).
bandIndex: Calculate the Chern number of the bandIndex band.
b1b2points: The number of points along the direction of two reciprocal lattice vectors. The default value is {50, 50}.
b1b2ranges: The range of points along the direction of two reciprocal lattice vectors. The default value is {2,2}.
	
Output:
KPoints: The coordinates of points in reciprocal space.
kPointFigure: K-point schematic diagram.
ChernNumber: Chern number.
BerryCurvatureData: Berry curvature data.
BerryCurvatureFigure3D: Plot with the ListPlot3D function.
BerryCurvatureFigureContour: Plot with the ListContourPlot function.
BerryCurvatureFigureDensity: Plot with the ListDensityPlot function.
	
Example:
h[{kx_,ky_,kz_},M_:0.1,t1_:1,t2_:0.2,\[Phi]_:0.2*Pi]=H;
lat={{-Sqrt[3],0,0},{Sqrt[3]/2,-3/2,0},{0,0,1}};
bnd=Table[i,{i,1,1}];
ChernNumberCalc[h,lat,bnd]
\end{lstlisting}

	\section{The relevant code input for calculating the $\mathbf{k} \cdot \mathbf{p}$ model}
	
	\subsection{Consider the $\mathbf{k} \cdot \mathbf{p}$ model with Rashba SOC}
	\label{D1}
	\begin{lstlisting}
H={{(kx^2+ky^2)/(2m),I \[Alpha](kx-I ky)},{-I \[Alpha](kx+I ky),(kx^2+ky^2)/(2m)}};
	
Clear[h]
h[{kx_,ky_,kz_},m_:1,\[Alpha]_:0.8]=H;
lat={{1,0,0},{0,1,0},{0,0,1}};
ds=HCalc2D[h,lat,{1,2},{50,50},{1,1},2,1,0];
Normal[ds[4]]
	
Clear[h]
h[{kx_,ky_,kz_},m_:1,\[Alpha]_:0.8]=H;
lat={{1,0,0},{0,1,0},{0,0,3}};
ds1=SpinTextureAtEnergy2D[h,lat,1,1,Red,1,0.006,1];
ds2=SpinTextureAtEnergy2D[h,lat,1,2,Blue,1,0.006,1];
Show[ds1[2],ds2[2]]
	\end{lstlisting}
	
	\section{The relevant code input for calculating the post-processing capabilities}
	
	\subsection{VASP data processing}
	\label{E1}
	\begin{lstlisting}
(*Import*)
path=SetDirectory[FileNameTake[NotebookFileName[],{1,-2}]];
poscar=Import["POSCAR","Table"];
kpoints=Import["KPOINTS","Table"];
procar=Import["PROCAR","Table"];
efermi=-2.0737;
	
(*ISPIN=1, UpOrDown=1 (any value is acceptable when ISPIN=1), SOC=2*)
(*Sz Spin Component*)
ds=VASPPBand[poscar,kpoints,procar,efermi,1,1,{2,3},{0},"PBand-Sz.dat"]
	
(*Extract data and plot*)
KK=Normal[ds[5]];
KK=KK/.{KK[[4,2]]->"-K"}
tpband=Normal[ds[7]];
Normal[ds[6]]
	
(*tot orbitals corresponds to 19 columns*)
(*Sz Spin Component*)
PBandSpinPlot[tpband,KK,{19},"\!\(\*SubscriptBox[\(S\), \(z\)]\)",3/4,{-1.2,0.7}]
	\end{lstlisting}	
	
	
	
	
	

	
	
	
	
	

\end{document}